\documentclass[prb, twocolumn, showpacs, amsmath, amssymb,superscriptaddress]{revtex4-1}
\usepackage{amssymb,amsfonts,amsmath} 
\usepackage{graphicx}
\usepackage[usenames,dvipsnames]{color}
\usepackage{xcolor}
\usepackage[colorlinks=true, citecolor=blue, linkcolor=blue, urlcolor=blue, breaklinks=True]{hyperref}
\usepackage{longtable}
\usepackage{bbm}
\usepackage{enumitem}
\usepackage{bm}
\usepackage{braket}
\usepackage{comment}
\usepackage{breakurl}
\usepackage{nicefrac}

\begin{document}
\title{Vortex control in superconducting Corbino geometry networks
}
\author{T. Okugawa}
\affiliation{Institut f\"ur Theorie der Statistischen Physik, RWTH Aachen, 
52056 Aachen, Germany and JARA - Fundamentals of Future Information Technology.}
\author{S. Park}
\affiliation{Departamento de Física Teórica de la Materia Condensada, Condensed Matter Physics Center (IFIMAC) and Instituto Nicolás Cabrera, Universidad Autónoma de Madrid, 28049 Madrid, Spain}
\author{P. Recher}
\email{p.recher@tu-braunschweig.de}
\affiliation{Institut f\"ur Mathematische Physik, Technische Universit\"at Braunschweig, D-38106 Braunschweig, Germany.}
\affiliation{Laboratory for Emerging Nanometrology Braunschweig, D-38106 Braunschweig, Germany.}
\author{D. M. Kennes}
\email{Dante.Kennes@rwth-aachen.de}
\affiliation{Institut f\"ur Theorie der Statistischen Physik, RWTH Aachen, 
52056 Aachen, Germany and JARA - Fundamentals of Future Information Technology.}
\affiliation{Max Planck Institute for the Structure and Dynamics of Matter, Center for Free Electron Laser Science, 22761 Hamburg, Germany.}

\date{\today}


\begin{abstract}
{In superconductors, vortices induced by a magnetic field are nucleated randomly due to some random fluctuations or pinned by impurities or boundaries, impeding the development of vortex based quantum devices. Here, we propose a superconducting structure which allows to nucleate and control vortices on-demand by controlling magnetic fields and currents. Using time-dependent Ginzburg Landau theory, we study a driven vortex motion in two-dimensional Corbino geometries of superconductor-normal metal-superconductor Josephson junctions. We remedy the randomness of nucleation by introducing normal conducting rails to the Corbino disk to guide the nucleation process and motion of vortices towards the junction. We elaborate on the consequences of rail-vortex and vortex-vortex interactions to the quantization of resistance across the junction. Finally, we simulate the nucleations and manipulations of two and four vortices in Corbino networks, and discuss its application to Majorana zero mode braiding operations. Our study provides a potential route towards quantum computation with non-Abelian anyons.}
\end{abstract}
\maketitle
\section{Introduction}
\label{introduction}
The key feature of type II superconductors is the anomalous mixed state where in the presence of magnetic fields normal conducting regions surrounded by persistent circular super-currents --- i.e. vortices --- nucleate in the superconducting material and the motion of these vortices without pinning causes electric  dissipation.~\cite{Bardeen1957, Tinkham, Abrikosov1957} It is well-known that most of the high-Tc superconductors are categorized as type II superconductors, and controlling vortices on demand is one of the important prerequisites for creating novel superconducting devices.~\cite{Blatter1994, 2004Wordenweber, 2020Jones,RevModPhys.93.041002}

In addition, there are some proposals regarding the emergence of non-trivial states based on type II superconducting hybrid systems where controllability of vortices is one of the important features.~\cite{Hals2016, Berciu_2005, Weeks_2007,RevModPhys.93.041002}
Moreover, recent studies proposed that a Majorana zero mode emerges at the core of vortices in topological superconductors, and a spatial exchange of these Majorana zero modes (we call it braiding operation) obey non-Abelian anyonic statistics where particle exchanges result in non-trivial operations which do not commute in general. Such non-commutative braids can be used as the basis for topological quantum computation.~\cite{Xian2011, Leijnse_2012, Elliott2015, Nayak2008, Cimento2017}  
In this regard, many theoretical proposals have been put forward both for realizing Majorana zero modes~\cite{Kane2008, Hell2017, Nadj2013,Choy2011,Alice2010, Oreg2010, Lutchyn2010, Read2000, Kitaev2001} and for performing Majorana braiding operations,~\cite{Alicea_2011, Mi2013, van_Heck_2012, Flensberg2011, Ivanov2001}. Despite all these efforts, an experimental demonstration of the braiding operation has not been realized yet, although signs of Majorana zero modes have already been detected in several experimental platforms.~\cite{Xu2015, Sun_2017, Wiedenmann_2016, Rokhinson_2012, Feldman_2016, Pawlak_2016, Lee_2013, Das_2012, Deng1557, Mourik1003, Deng2012, Deacon2017}

A promising and natural set-up to perform braiding operations is a Corbino geometry superconducting-normal-superconducting topological Josephson junction in a superconducting thin film, 
where a circular junction is located between two superconducting electrodes and an external current flows inwards/outwards through the Josephson junction.
As a consequence, the vortices in the Corbino geometry should be trapped on the circular junction and execute a circular motion along the Josephson junction due to the Lorentz force. There are already some theoretical studies of such specific set-ups for the purpose of Majorana braiding and also a few experimental investigations.~\cite{2010Clem, Sunghun2015, Sunghun2020, Grosfeld2011, Hadfield2003, Matsuo2020}              


Another crucial question is how vortices enter the circular junction when a homogeneous perpendicular magnetic field is applied to the structure. In a perfect circular junction, vortices are created by spontaneous symmetry breaking (which breaks the rotation symmetry of the circle) without control of the exact nucleation position. Here, we address the nucleation of several vortices on a circular junction as needed for braiding of vortices and how an enhanced control over vortex dynamics can be achieved via rails guiding them into the circular junction. We also show that the resistance across the Corbino geometry Josephson junction develops jumps associated with the entrance of single vortices. 

Although there are already some theoretical studies on Corbino geometry Josephson junctions, most of them employ the microscopic BCS-model and there are only a few studies focusing on macroscopic vortex dynamics. 
In this paper, we explore vortex dynamics in Corbino geometry Josephson junctions in the presence of both an external current and an applied magnetic field using phenomenological time-dependent Ginzburg Landau (TDGL) theory.~\cite{Gorkov1959, Schmid1966, Cyrot1973} 
Here, we focus on the simplest case of non-topological $s$-wave superconductors, however, the main conclusions of our study should carry over to the case of topological ($p-$ or $d-$wave) superconductors, where Majorana bound state resides at the core of vortices since the vortex dynamics is expected to be qualitatively similar regardless of the type of pairing.
At the end of this paper, we show simulations of braiding for two vortices in real time by controlling an external current and an applied magnetic field. Furthermore, we also demonstrate braiding operations for two vortices in the presence of two other vortices in the superconducting sample where the exchange statistics of non-Abelian anyons can be utilized for protected quantum operations. We believe that our proposed set-ups and simulations of vortex dynamics deliver a platform for scalable registers of Majorana based qubits with read-in/read-out as well as braiding capabilities.

Our focus is on the vortex motion in connection with Corbino geometry Josephson junctions, however, there is no reason to believe that general features shown in this paper with respect to vortex-control cannot be observed and employed also in different type II superconductor based hybrid systems.

The rest of this paper is organized as follows. In section \ref{formalism}, we present our model and the computational method adopted for this paper. In section \ref{result}, we show the main results of our calculations and the corresponding discussion. Finally, section \ref{conclusion} concludes this paper.

\section{Formalism}
\label{formalism}
We employ TDGL equations to simulate the dynamics of a complex superconducting order parameter $\Delta=|\Delta|e^{\mathrm{i}\phi}$ in a two dimensional disk geometry in the presence of both an external magnetic field $\bm{B}$ directed perpendicular to the two-dimensional superconductor and an external current created by a source and drain of particles whose strength is denoted by $Q$. The electromagnetic field is represented by the vector potential $\bm{A}$ and the scalar potential $\Theta$. Denoting the charge density as $\rho$ and current density as $\bm{J}$ and taking $\hbar=c=e=1$ as units, the equations are given as follows:~\cite{Tinkham, Kennes, Benyamini_2019, Benyamini_2020}
\begin{align}
&\frac{1}{D} \left(\partial_{t} +2\mathrm{i} \Psi \right)\Delta = \frac{1}{\xi^2 \beta} \Delta [\alpha(\bm{r}) - \beta |\Delta |^2] \notag \\
&+ [ \bm{ \nabla} - 2 \mathrm{i}\bm{A}  ]^2 \Delta,  \label{TDGL} \\
&\bm{J}=\sigma [ -\bm{\nabla}\Psi - \partial_{t}\bm{A}  ] + \sigma \tau_{s} \Re[\Delta^{\ast} \left( -  \mathrm{i} \bm{\nabla} - 2 \bm{A}      \right)\Delta ], \label{current} \\
&\rho = \frac{\Psi - \Theta}{4 \pi \lambda_{TF}^2}, \label{final} \\
&\partial_{t} \rho + \bm{\nabla} \cdot \bm{J}=Q, \label{continuity}\\
&\bm{\nabla}^2 \Theta = -4 \pi \rho, \label{poisson}
\end{align}
where $D$ is the normal state diffusion constant, $\Psi$ is the electrochemical potential per electron charge, $\xi=\sqrt{6D/\tau_{s}}$, and the superconducting coherent length is given as $\xi_{0} = \xi/\sqrt{\alpha(\bm{r})/\beta}$. $\tau_{s}$ is the spin-flip scattering time. $\beta$ is a system dependent constant, which sets the magnitude of the order parameter. $\alpha(\bm{r}) \propto  [T_c-T] $ is the spatially dependent parameter which governs the  superconducting or normal state at $\bm{r}$ and 
where $T_{(c)}$ is the (critical) temperature. $\sigma$ and $\lambda_{TF}$ are the normal state conductivity and the Thomas-Fermi static charge screening length, respectively. 
We measure length in units of $\xi$ and time in units $\xi^2/D$ (since $\xi$ is the unit of length, we write this as $D^{-1}$). The parameters are chosen as $\beta=1$, $\tau_s=6 D/\xi^2$, $\lambda_{TF}/\xi=1$, and $\sigma/(D/\xi^2)=1$ and we employ the coulomb gauge $\nabla\cdot A=0$.~\footnote{However, this condition does not uniquely fix the Gauge function $\chi$ for finite element methods used here, i.e $\chi$ can only be fixed up to a constant. Thus, we impose another condition; $\int_S \Psi dS=0$, where $S$ is the entire sample domain.~\cite{Qiang1994, fan2019}}

The radius of the finite disk geometry is set to be $17 \xi$. We chose these parameters for definiteness, but verified that none of the general conclusions is affected by this particular choice. To model Corbino geometry superconducting-normal-superconducting Josephson junction, the $\alpha(\bm{r})$ parameter is set accordingly, namely $\alpha>0$ and $\alpha<0$ in superconducting and normal  region, respectively. We define the circular normal metallic region to be on a disk with radius $6.8\xi$ and width $0.5\xi$ and assign $\alpha=-1$ while the other region in the sample is superconducting with $\alpha=1$. We refer to this as a normal Corbino geometry set-up and later in this paper, we introduce metallic regions in addition to the normal Corbino geometry to improve the control of vortices (see section \ref{result}).         

The simulations are performed using a finite element method in space implemented in FEniCS.~\cite{LoggMardalEtAl2012a} As is a standard approach to the time-dependent problem within our finite element method, we discretize the time derivative by a finite difference approximation. Specifically, the discritization used here is in steps of $D\Delta t = 0.5$. This allows to obtain numerically converged results which was verified by decreasing this numerical parameter and checking that the results remain unaffected.\footnote{There is only one exception and that is the case of the normal Corbino geometry set-up without any impurities or rails (see below) where a slight change of $D\Delta t$ causes the nucleation points of vortices and the number of vortices to be different as they are determined by random numerical fluctuations. This is discussed further in section \ref{conclusion}.
} In the calculations, we impose superconductor-vacuum boundary condition along the outer circle edge of the finite sample:
\begin{align}
\left(-\mathrm{i} \bm{\nabla}\Delta - 2\bm{A}\Delta \right)\cdot \bm{n}  &= 0~~\text{on boundary},   \label{no_bound_current}\\
\left(-\bm{\nabla}\Psi - \partial_{t}\bm{A}  \right)\cdot \bm{n}  &= 0~~\text{on boundary},  \label{no_normal_current}
\end{align}
where $\bm{n}$ is the unit vector normal to the boundary. 
To integrate the partial deferential equation, initial conditions at time $t=0$ are needed which we set to zero for all variables except $\Delta$,  while we choose $\Delta(x, y, t=0)$ to be in the superconducting state with some random fluctuation over the entire 2D sample. \footnote{Specifically, both real and imaginary part of the complex order parameter $\Re[\Delta]$ and $\Im[\Delta]$ are set to be $\sqrt{0.5}+\delta$, where $\delta$ is chosen randomly from a uniform distribution $\delta \in (-0.01, 0.01]$. However, since we concentrate on analyzing the steady state, none of the general conclusions are affected by the initial conditions.}  

We consider very thin films for which the London penetration depth is much larger than the thickness of the sample so that the magnetic field can be taken to be spatially uniform and described by the Coulomb-gauged vector potential $\bm{A}=(-By, Bx, 0)/2$, simplifying the boundary condition, Eq. \eqref{no_bound_current}\eqref{no_normal_current} further.

The external current is introduced by defining source and drain sections at the vicinity of the open boundary of the superconducting disk and that of the center, respectively (see Appendix \ref{appendix:source} for details). The width of the source region is $0.5\xi$. 
The drain section is set to be a disk whose radius is $2.9\xi$ as shown in FIG. \ref{fig:ju} in Appendix \ref{appendix:source}. Thus, the current flow is homogeneous in almost the entire sample (except for the source and drain sections). The current flows from the circle boundary towards the center of the sample.    

\section{Results}
\label{result}
\subsection{Basis of vortex control}
\begin{figure}[htp!]
\begin{center} 
\includegraphics[width=0.5\textwidth, angle=-0]{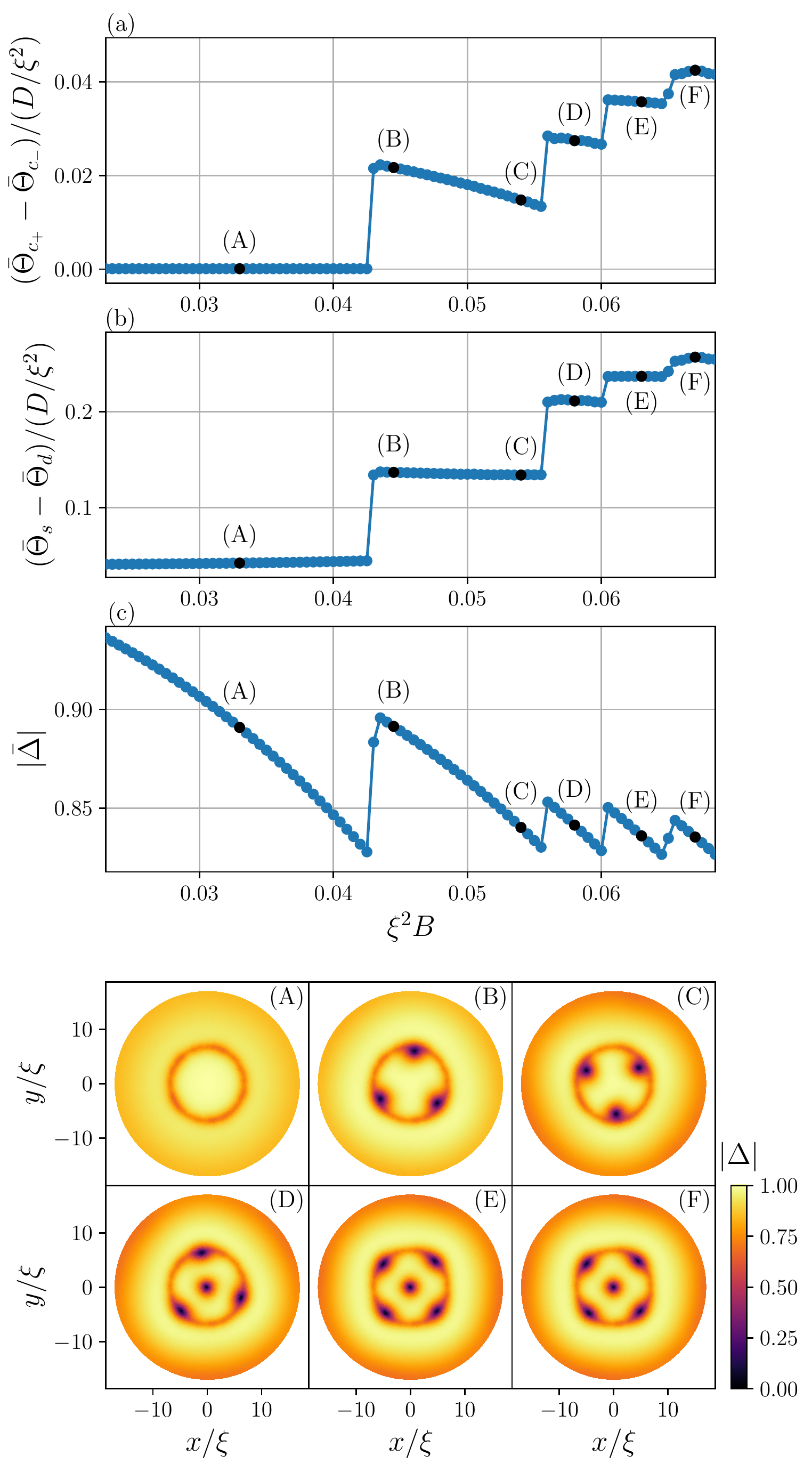}
\caption{Upper panels (a)(b) and (c) show the resistance  $\Delta \bar \Theta$ and the time-space-average of the absolute value of the order parameter $|\bar \Delta|$ as a function of the applied field $B$, respectively. Here, we calculate the resistance by taking differences of the averaged scalar potential $\Bar{\Theta}$ between (a) two nearby points across the Josephson junction, denoted by $\Bar{\Theta}_{c_{+/-}}$ and (b) source and drain points, denoted by $\Bar{\Theta}_{s/d}$. The average is taken over both time and space. 
For the scalar potential $\bar \Theta$ in panel (a) and (b), the space-average is taken over $100$ central angles from $0$ to $2\pi$ 
while the order parameter$|\bar \Delta|$ in panel (c) is calculated by averaging over all $|\Delta|$ values in the entire sample. The time-average is taken over $200$ time steps for both $|\Delta|$ and $\Theta$ after the space-average has been performed. The applied $B$ field is increased in steps of $5 \times 10^{-4} \xi^{-2}$ every 400 time steps from which the second $200$ time steps are used for the time-averaging 
when the stationary vortex motions have been established. The strength of the source term is chosen to be $Q/(D^2/\xi^6)=0.2$. Lower panel shows snapshots of the absolute value of the order parameter $|\Delta|$ at the labeled points (A)-(F) in the upper panel (a)(b)(c). The corresponding video is available in the SM\cite{SM} (see also TABLE \ref{table:video} in Appendix \ref{appendix:video}).} 
\label{fig:corbino_snap_resistance}
\end{center}
\end{figure}
In order to connect our simulation results to experimentally accessible quantities, we consider the resistances $R=\Delta \bar \Theta/Q$, where $\Delta \bar \Theta$ is the difference of the scalar potential averaged in time and radially between two measurement points. Since the externally sourced current is constant in our system (except in the source and drain sections), we can consider differences of the scalar potential $\Delta \bar \Theta$ to be directly proportional to the resistance (thus, we denote $\Delta \bar \Theta$ as resistance hereafter). Specifically, in this paper we calculate $\Delta \bar \Theta$ from the source and drain points, denoted by $\Bar{\Theta}_{s}-\Bar{\Theta}_{d}$ and two nearby points across the Josephson junction, denoted by $\Bar{\Theta}_{c_{+}}-\Bar{\Theta}_{c_{-}}$.

First, we show the results of the normal Corbino geometry set-up, which is summarized in FIG. \ref{fig:corbino_snap_resistance}, where we plot the averaged resistances  
$\Bar{\Theta}_{s/c_{+}}-\Bar{\Theta}_{d/c_{-}}$ in (a),(b) and in (c) the absolute value of the order parameter $\Bar{|\Delta|}$ as a function of the applied magnetic field $B$ (upper panel) and snapshots of the absolute value of the order parameter in the lower panel. 
The upper panel shows successive resistance plateaus with jumps in between [(a),(b)] while in (c) $\Bar{|\Delta|}$ displays an almost linear decrease as $B$ increases with jumps at the same points as the plateau transitions in (a),(b). These jumps coincide with an increase in the number of vortices as shown in the lower panel (see the SM video1~\cite{SM}). For the first plateau in (a),(b) and linear slope in (c) of the upper panel, there is no corresponding vortex in the sample as shown in the snapshot (A) of the lower panel. 
After this, a sudden jump is observed in both the resistances $\Bar{\Theta}_{s/c_{+}}-\Bar{\Theta}_{d/c_{-}}$ and the absolute value of the order parameter $\Bar{|\Delta|}$ at around  $B\xi^2=0.0425$ in (a)(b) and (c) of the upper panel. This signifies a nucleation of vortices, followed by a circular stationary motion along the Josephson junction due to the Lorentz force created by the inward sourced current; compare the snapshots of (A) and (B) in the lower panel. 
Then, the first non-zero resistance plateau in (a),(b) and linear decrease of $\Bar{|\Delta|}$ in (c) of the upper panel follow, and the vortex configuration stays the same as shown in the snapshots (B) and (C).~\footnote{Here, the resistance measured in the vicinity of the Josephson junction shows a small decrease, which is due to the shift of vortices trapped on the Josephson junction toward the center 
as can be seen from the comparison of the snapshots (B) and (C) in the lower panel. This slight shift causes the small decrease in the resistance measured in the vicinity of the Josephson junction because the two nearby measuring points across the Josephson junction are fixed while vortices are pushed toward the center. Comparing to other plateaus in panel (a) of FIG. \ref{fig:corbino_snap_resistance}, this is the only plateau which shows a slight decrease. This is because there is one special vortex residing at the center of the superconducting sample 
which prevents other vortices on the Josephson junction from being pushed toward the center due to vortex-vortex interactions.} After this, another vortex nucleates as shown in the snapshot (D) of the lower panel, and a new resistance plateau shown in panel (a),(b) and a new linear slope in (c) develop. 
This correspondence among (a),(b) and (c) in the upper panel and the snapshots of $|\Delta|$ in the lower panel holds even with increasing $B$-field. The height of the resistance plateau in (a) and (b) are almost the same among the last 3 resistance plateaus as expected as newly nucleated vortices are always trapped on the Josephson junction. 
Here, one thing to note is that the height differences between the first to second and the second to third of the resistance plateaus in (b) of the upper panel is almost the same even though the number of vortices change from $0$ to $3$ and $3$ to $4$, respectively. This is due to the existence of a vortex at the center of the sample, which gives rise to a deep downward slope toward the center in $\bar \Theta$ (see FIG \ref{fig:theta_normal} in the Appendix \ref{appendix:scalar} for details), which results in relatively large jumps in the resistance measured from the source and drain points $\Bar{\Theta}_{s}-\Bar{\Theta}_{d}$ in (b) of the upper panel. This can also be understood from the fact that the height differences between the second to third and the third to fourth resistance plateau in (b) of the upper panel are very different and the latter is almost $3$ times as small as the one of the first to second, which corresponds to an increase in the number of vortices from $0$ in the snapshot (A) to $3$ in (B) and $4$ in (D) to $5$ in (E). 

Overall, we identify mainly three issues in the normal Corbino geometry set-up as far as controlling vortex motion is concerned. First, the number of vortices does not increase one by one as is seen by comparing the snapshots of (A) and (B) in the lower panel. Second, nucleation points of the vortices cannot be controlled.~\footnote{When a vortex nucleates into the sample, the rotational symmetry of the Corbino geometry is spontaneously broken.~\cite{2010Clem, Sunghun2015, Sunghun2020} Tiny inhomogeneities such as defects, thermal fluctuations, or even numerical flucatuations matter for the system to decide where the vortices nucleate.} 
Third, vortices also go into the center of the sample rather than being trapped on the Josephson junction as shown in (C)-(F) of the lower panel. Thus, the number of vortices to nucleate and spatial points of nucleation are out of control.~\footnote{In the case of the numerical simulation, we observe that even a slight modification of the meshing configuration required for the finite element method changes the nucleation points and the number of trapped vortices, when the system starts to develop vortices, which is also reported in Ref.~\onlinecite{Pack2020}.} However, the vortices that are trapped in the circular Josephson junction are equally spaced on the circular junction, which is consistent with findings in Refs.~\onlinecite{2010Clem, Sunghun2015, Sunghun2020}

\begin{figure}[t!]
\begin{center} 
\includegraphics[width=0.5\textwidth, angle=-0]{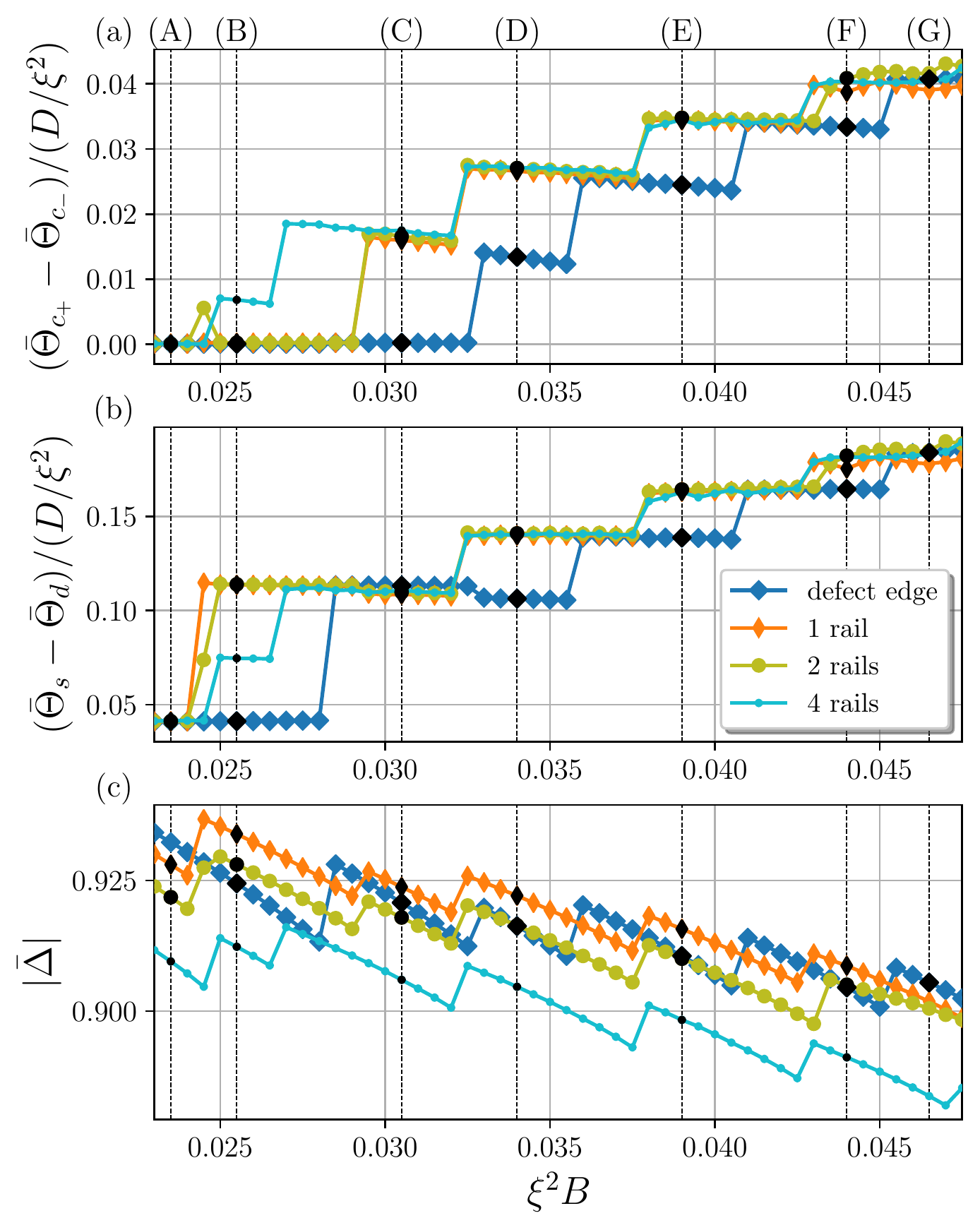}
\caption{Comparison of the the resistances $\Delta \bar \Theta$ in (a),(b) and (c) a spatio-temporal average of the absolute value of the order parameter $\Bar{|\Delta|}$ among the different set-ups (compare FIG. \ref{fig:defect1_snap} and \ref{fig:24rail_snap} for defect edge/1 rail and  2/4 rails set-ups, respectively) as a function of the applied $B$ field. The resistance is calculated by taking differences of the averaged scalar potential $\Bar{\Theta}$ between (a) two nearby points across the Josephson junction, denoted by $\Bar{\Theta}_{c_{+/-}}$ and (b) source and drain points, denoted by $\Bar{\Theta}_{s/d}$ (see Appendix \ref{appendix:scalar} for details). The same averaging procedure and a strength of the source term $Q/(D^2/\xi^6)$ are used as in FIG. \ref{fig:corbino_snap_resistance}. For the black data points labeled (A)-(G) at the top of the figure (black dotted vertical lines in the figure are guide for the eye), the corresponding snapshots of the absolute value of the order parameter $|\Delta|$ for each set-up are shown in FIG. \ref{fig:defect1_snap} and \ref{fig:24rail_snap}. All corresponding videos are available in the SM\cite{SM} (see also TABLE \ref{table:video} in Appendix \ref{appendix:video}).}
\label{fig:comp_all_0124N}
\end{center}
\end{figure}%

\begin{figure}[t!]
\begin{center} 
\includegraphics[width=0.5\textwidth, angle=-0]{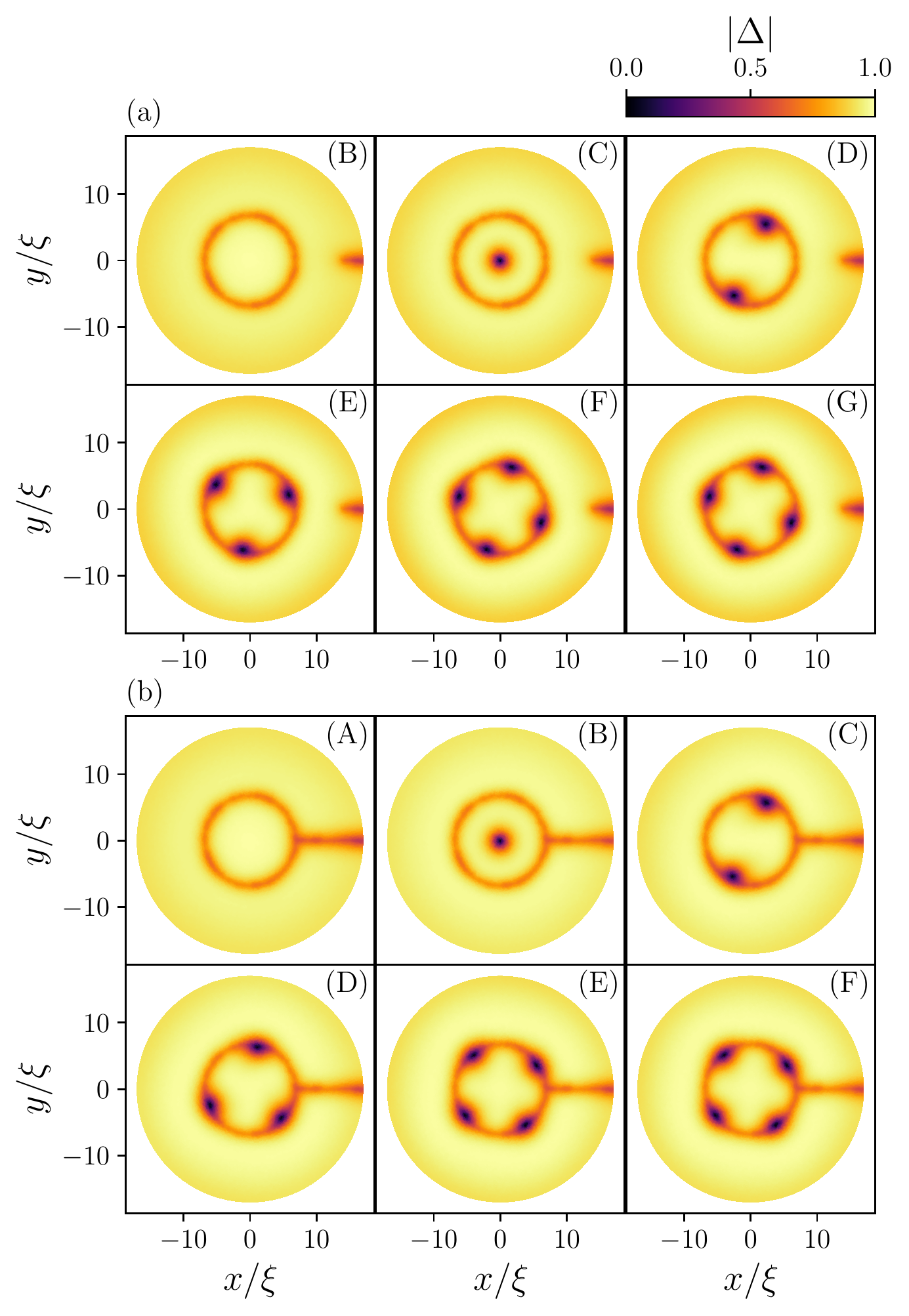}
\caption{Snapshots of the absolute value of the order parameter $|\Delta|$ at the black points in the plots of FIG. \ref{fig:comp_all_0124N} for the defect edge set-up (a) and the 1 rail set-up (b). All corresponding videos are available in the SM\cite{SM} (see also TABLE \ref{table:video} in Appendix \ref{appendix:video}).}
\label{fig:defect1_snap}
\end{center}
\end{figure}
\begin{figure}[t]
\begin{center} 
\includegraphics[width=0.5\textwidth, angle=-0]{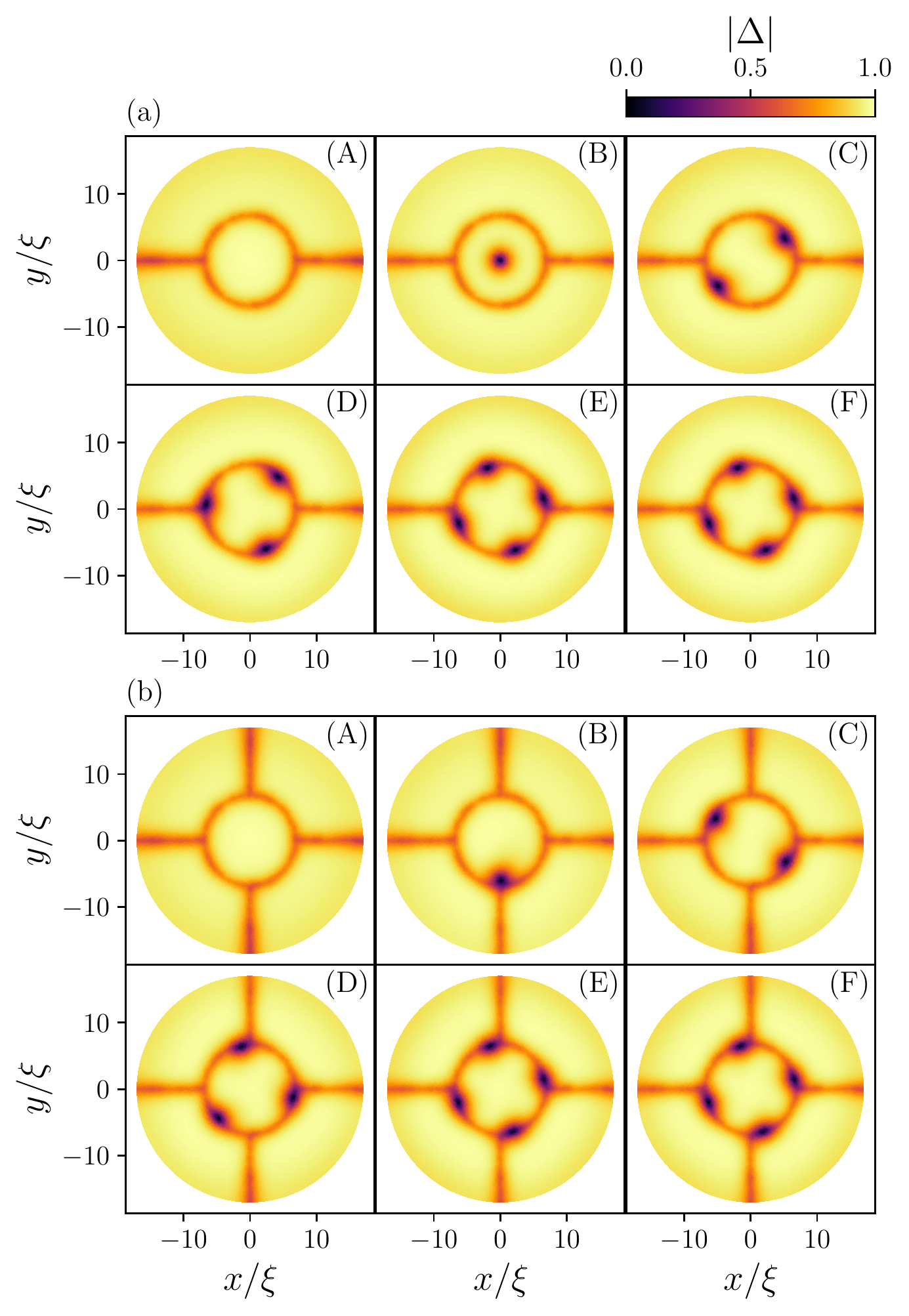}
\caption{Snapshots of the absolute value of the order parameter $|\Delta|$ at the black points in the plots of FIG. \ref{fig:comp_all_0124N} for the 2 rails set-up (a) and the 4 rails set-up (b). All corresponding videos are available in the SM\cite{SM} (see also TABLE \ref{table:video} in Appendix \ref{appendix:video})}
\label{fig:24rail_snap}
\end{center}
\end{figure}

In order to avoid the three issues above, we add other metallic regions to the normal Corbino geometry set-up to stabilize vortex nucleation. In this paper, we introduce narrow metallic lines connected to the outside of the superconducting film, with the suggested geometries shown in FIG. \ref{fig:defect1_snap} and \ref{fig:24rail_snap}. We call them as follows: the set-up of FIG. \ref{fig:defect1_snap}(a) "defect edge", FIG. \ref{fig:defect1_snap}(b) "1 rail", FIG. \ref{fig:24rail_snap}(a) "2 rails", and FIG. \ref{fig:24rail_snap}(b) "4 rails". We perform the same analysis as in FIG. \ref{fig:corbino_snap_resistance} with these new set-ups and compare the pros and cons with respect to being able to control vortices. The summarized results and the corresponding snapshots of the absolute value of the order parameter for each set-up are shown in FIG. \ref{fig:comp_all_0124N}, \ref{fig:defect1_snap} and \ref{fig:24rail_snap}. First, in the case of the defect edge set-up, we observe that vortices nucleate one by one at the defect point and nucleation coincides with jumps of the average of the absolute value of the order parameter $\Bar{|\Delta|}$ (see the blue line in panel (c) of FIG. \ref{fig:comp_all_0124N} and panel (a) of FIG. \ref{fig:defect1_snap}). However, the first nucleated vortex goes into the center of the sample rather than being trapped on the Josephson junction as can be seen in snapshot (C) in panel (a) of FIG. \ref{fig:defect1_snap}. This vortex configuration can be also understood by comparing the results from the two types of resistance measurements, namely there is no change in $\Bar{\Theta}_{c_{+}}-\Bar{\Theta}_{c_{-}}$ while there is a jump in $\Bar{\Theta}_{s}-\Bar{\Theta}_{d}$ (see the blue line up to the point (C) in panel (a) and (b) of FIG. \ref{fig:comp_all_0124N}). From the points (C) to (D), another vortex is nucleated and two vortices are trapped on the Josephson junction executing their stationary circular motions. 
Then, further increase of $B$ induces one by one vortex nucleation and all vortices stay on the Josephson junction.

In the case of the 1 rail and 2 rails set-ups, again the first vortex cannot be trapped on the Josephson junction but goes into the center of the sample as shown in the snapshots of (B) in panel (b) of FIG. \ref{fig:defect1_snap} and panel (a) of FIG. \ref{fig:24rail_snap} although the 1 rail set-up shows one by one vortex nucleation. 
The resistance behavior of the 1 and 2 rail set-ups are almost identical with overlapping plots in panel (a),(b) of FIG. \ref{fig:comp_all_0124N} except for the final plateau in the magnetic field range from approximately $B\xi^2=0.043$ till the end, where the number of vortices change from $4$ to $5$ in the 1 rail set-up while it does change from $4$ to $6$ in the 2 rail set-up, see the snapshots (E) and (F) in panel (b) of FIG. \ref{fig:defect1_snap} and (a) of FIG. \ref{fig:24rail_snap}, respectively. Also, $\Bar{|\Delta|}$ is slightly smaller in the 2 rails set-up than 1 rail set-up due to an additional metallic railing in the 2 rails set-ups, which reduces the overall order parameter. 

\begin{figure}[t!]
\begin{center} 
\includegraphics[width=0.5\textwidth, angle=-0]{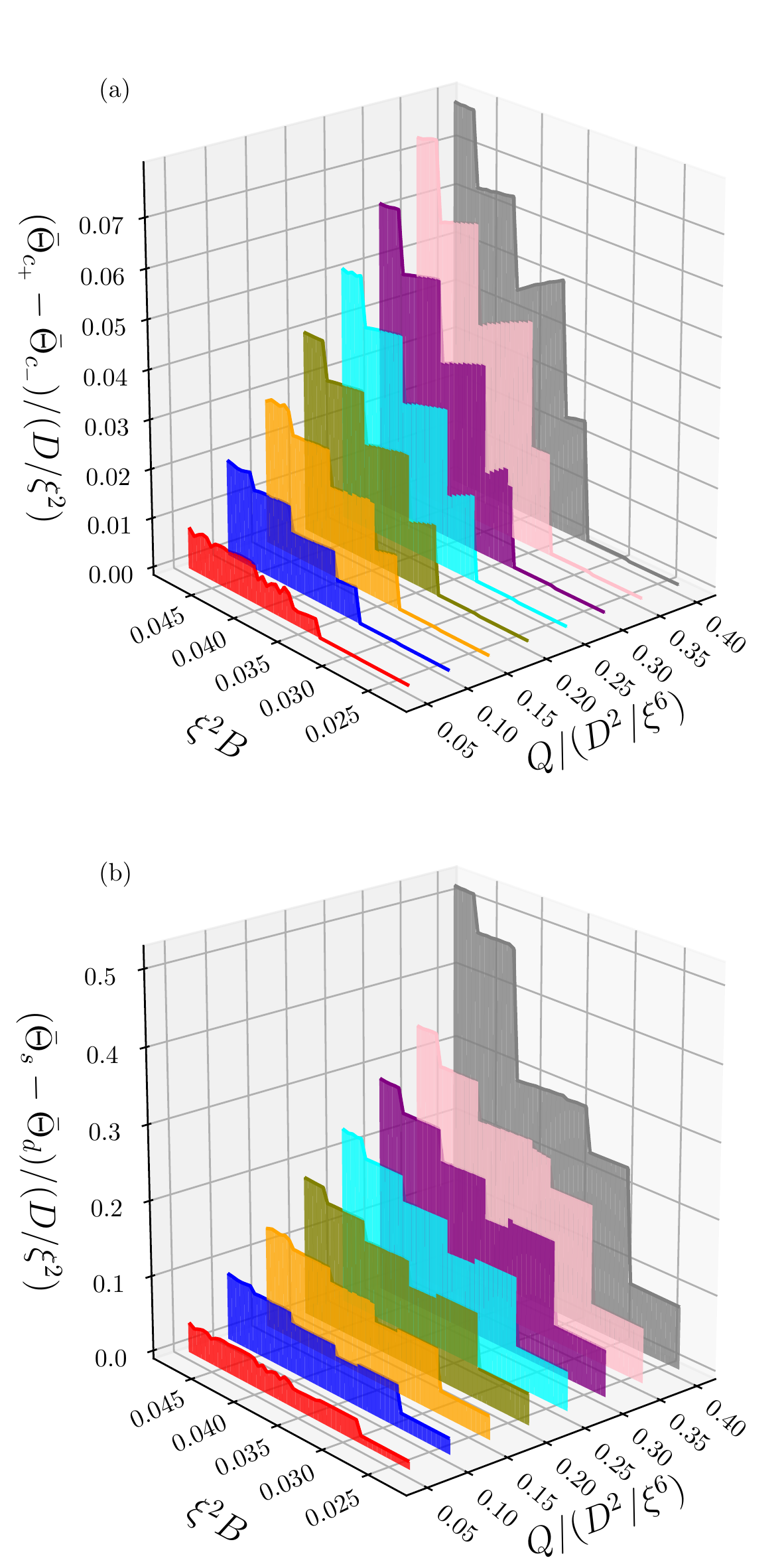}
\caption{Phase diagram of the defect edge set-up from the resistance $\Delta \bar \Theta$ as a function of strengths of the source term $Q$ and the applied field $B$, where the resistances are calculated by taking differences of the averaged scalar potential $\Bar{\Theta}$ between (a) two nearby points across the Josephson junction, denoted by $\Bar{\Theta}_{c_{+/-}}$ and (b) source and drain points, denoted by $\Bar{\Theta}_{s/d}$. The same averaging procedure is used as in FIG. \ref{fig:corbino_snap_resistance}. All corresponding videos are available in the SM\cite{SM} (see also TABLE \ref{table:video} in Appendix \ref{appendix:video}).}  
\label{fig:phase_defect}
\end{center}
\end{figure}

\begin{figure}[t!]
\begin{center} 
\includegraphics[width=0.5\textwidth, angle=-0]{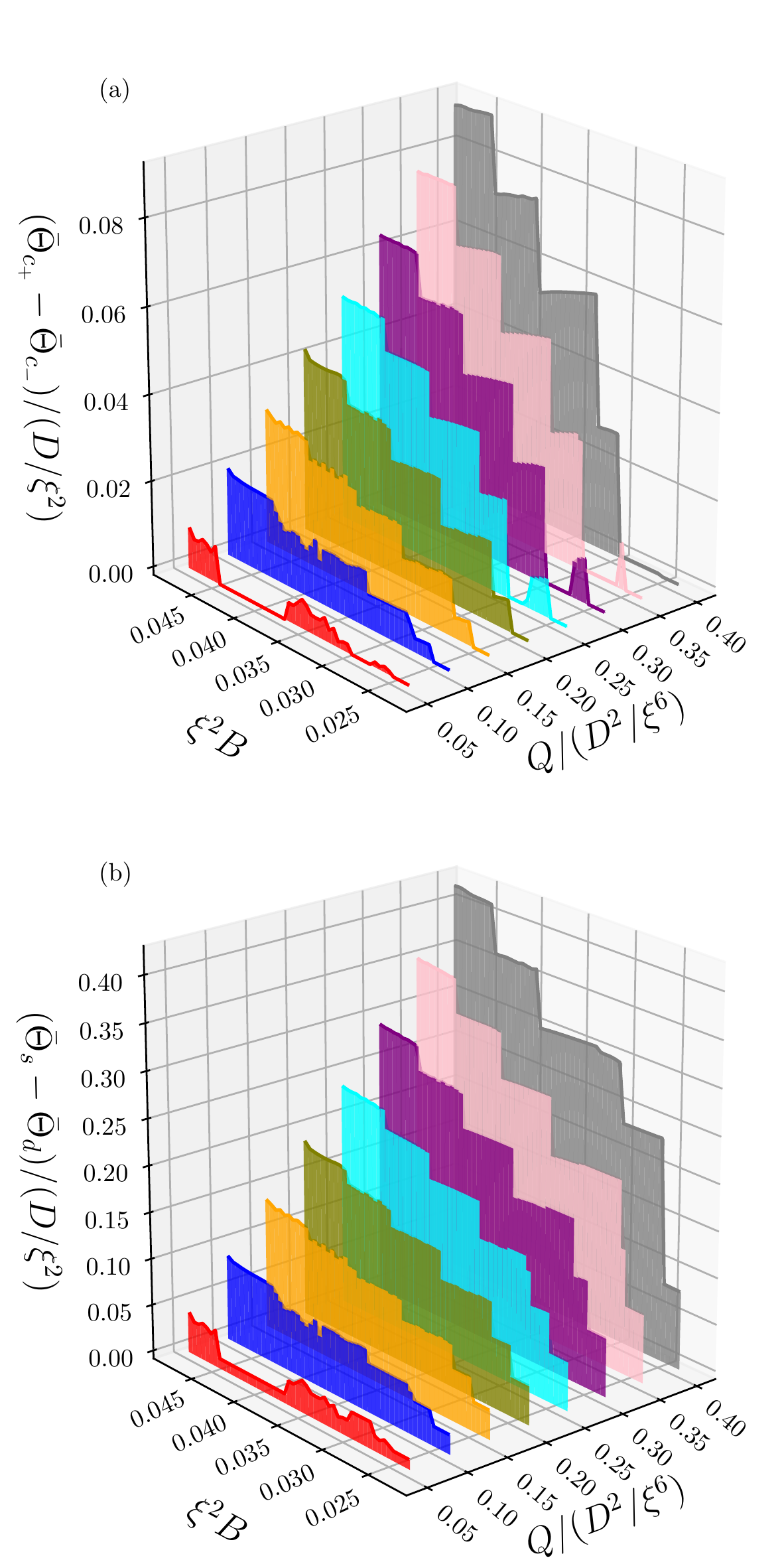}
\caption{Phase diagram of 4 rails set-up from the resistance $\Delta \bar \Theta$ as a function of strengths of the source term $Q$ and the applied field $B$, where the resistances are calculated by taking differences of the averaged scalar potential $\Bar{\Theta}$ between (a) two nearby points across the Josephson junction, denoted by $\Bar{\Theta}_{c_{+/-}}$ and (b) source and drain points, denoted by $\Bar{\Theta}_{s/d}$. The same averaging procedure is used as in FIG. \ref{fig:corbino_snap_resistance}. All corresponding videos are available in the SM\cite{SM} (see also TABLE \ref{table:video} in Appendix \ref{appendix:video}).} 
\label{fig:phase_4rail}
\end{center}
\end{figure}

Finally, unlike the other 3 set-ups, the 4 rails set-up shows perfect one by one nucleation, followed by a circular stationary vortex motion on the Josephson junction rather than vortices going into the center of the sample. This can be clearly seen from the snapshots of the absolute value of the order parameter in panel (b) of FIG. \ref{fig:24rail_snap}. 
Furthermore, the transition points in both resistances $\Bar{\Theta}_{s/c_{+}}-\Bar{\Theta}_{d/c_{-}}$ and the absolute value of the order parameter $\Bar{|\Delta|}$ perfectly coincide as seen in panel (a),(b) and (c) of FIG. \ref{fig:comp_all_0124N}, where each plateau in (a),(b) and each linear slope in (c) indicate different numbers of vortices in the system. Also, the fact that $\Bar{|\Delta|}$ is smaller as in the other set-ups is due to the additional metallic railings as already suggested for the comparison between 1 and 2 rails set-ups. Overall, the 4 rail set-up provides the best on-demand control of vortices among these four set-ups. 
One drawback of the railing set-ups (1,2, and 4 rails in this paper) is that the vortex motion becomes dragged by the intersections between the Josephson junction and the rails when they circulate around the Josephson junction, which disturbs a smooth stationary vortex motion on the Josephson junction and this effect does not arise in the case of the normal Corbino and defect edge set-up (compare the SM video1-video5~\cite{SM}).

Next, we further investigate the defect edge and the 4 rails set-ups by studying the dependence on the strength of the source term $Q$. Its effect on the resistance $\Bar{\Theta}_{s/c_{+}}-\Bar{\Theta}_{d/c_{-}}$ for the defect edge and the 4 rails set-ups are shown in FIG. \ref{fig:phase_defect} and \ref{fig:phase_4rail}, respectively. 

In the case of the defect edge set-up, as we have seen from the analysis of this set-up discussed in the FIG. \ref{fig:comp_all_0124N}, the first nucleated vortex goes into the center of the sample rather than being trapped on the Josephson junction for any value of the source term $Q$, which is clearly seen by comparing the resistance values in (a) and (b) of FIG. \ref{fig:phase_defect} where there is no resistance jump in $\Bar{\Theta}_{c_{+}}-\Bar{\Theta}_{c_{-}}$ shown in panel (a) while there is a relatively large jump in $\Bar{\Theta}_{s}-\Bar{\Theta}_{d}$ plotted in panel (b) at the magnetic field strength approximately $B\xi^2=0.030$. Another notable point is that there is a relatively large jump approximately $B\xi^2=0.0425$ for $Q/(D^2/\xi^6)=0.40$ in panel (b). This is because one of the vortices moves into the center of the sample when a new vortex is nucleated (see the SM video12~\cite{SM}), which causes the large jump in the resistance $\Bar{\Theta}_{s}-\Bar{\Theta}_{d}$. Comparison among different $Q$ values shows that a larger source current yields a larger resistance jump, as faster vortex motion induces a larger resistance~\cite{Tinkham, 2010Clem, Sunghun2015, Sunghun2020} while the jumps between different plateaus happen at almost the same $B\xi^2$-value regardless of the strength of $Q$.

For the 4 rails set-up in FIG. \ref{fig:phase_4rail}, in the region $0.10 \leq Q/(D^2/\xi^6) \leq 0.20$, the resistances  $\Bar{\Theta}_{s/c_{+}}-\Bar{\Theta}_{d/c_{-}}$ show a similar behavior for these $Q$-values. The first vortex is successfully trapped on the Josephson junction performing a stationary circular motion. For $0.25 \leq Q/(D^2/\xi^6) \leq 0.35$, although the resistance  $\Bar{\Theta}_{c_{+}}-\Bar{\Theta}_{c_{-}}$ shows a plateau between approximately $B\xi^2=0.025$ and $0.030$ (panel (a)), this plateau does not persist and the resistance vanishes again for larger $B$-fields. The resistance  $\Bar{\Theta}_{s}-\Bar{\Theta}_{d}$, however, shows another jump at around $B\xi^2=0.030$. This signifies that the first vortex stays on the Josephson junction from its nucleation time to around $B\xi^2=0.030$ and then moves into the center of the sample afterwards (see the SM video16-18~\cite{SM}). In the case $Q/(D^2/\xi^6)=0.40$, we can observe the same feature concerning the entrance of the first vortex as in the defect edge set-up, namely there is no signal in the resistance $\Bar{\Theta}_{c_{+}}-\Bar{\Theta}_{c_{-}}$ between approximately $B\xi^2=0.025$ and $0.030$ while there are large jumps in $\Bar{\Theta}_{s}-\Bar{\Theta}_{d}$, which indicates that the first vortex goes into the center of the sample rather than being trapped on the Josephson junction (see the SM video19~\cite{SM}). Furthermore, there appear resistance drops in $\Bar{\Theta}_{s/c_{+}}-\Bar{\Theta}_{d/c_{-}}$ between $B\xi^2=0.0375$ and $0.045$ for $Q/(D^2/\xi^6)=0.05$. 
Here, all four vortices in the system are stuck at each of the intersections of the Josephson junction with the rails because the source current is not large enough to drive all vortices pass the intersections against the attractive force and intricate vortex-vortex interaction (see the SM video13~\cite{SM}). 
Overall, if the strength of source term $Q$ is too small, the railing intersections strongly disturb the smooth stationary motion of vortices on the Josephson junction whereas too large currents fail to trap the first vortex on the Josephson junction.                    

\begin{figure}[t!]
\begin{center} 
\includegraphics[width=0.5\textwidth, angle=-0]{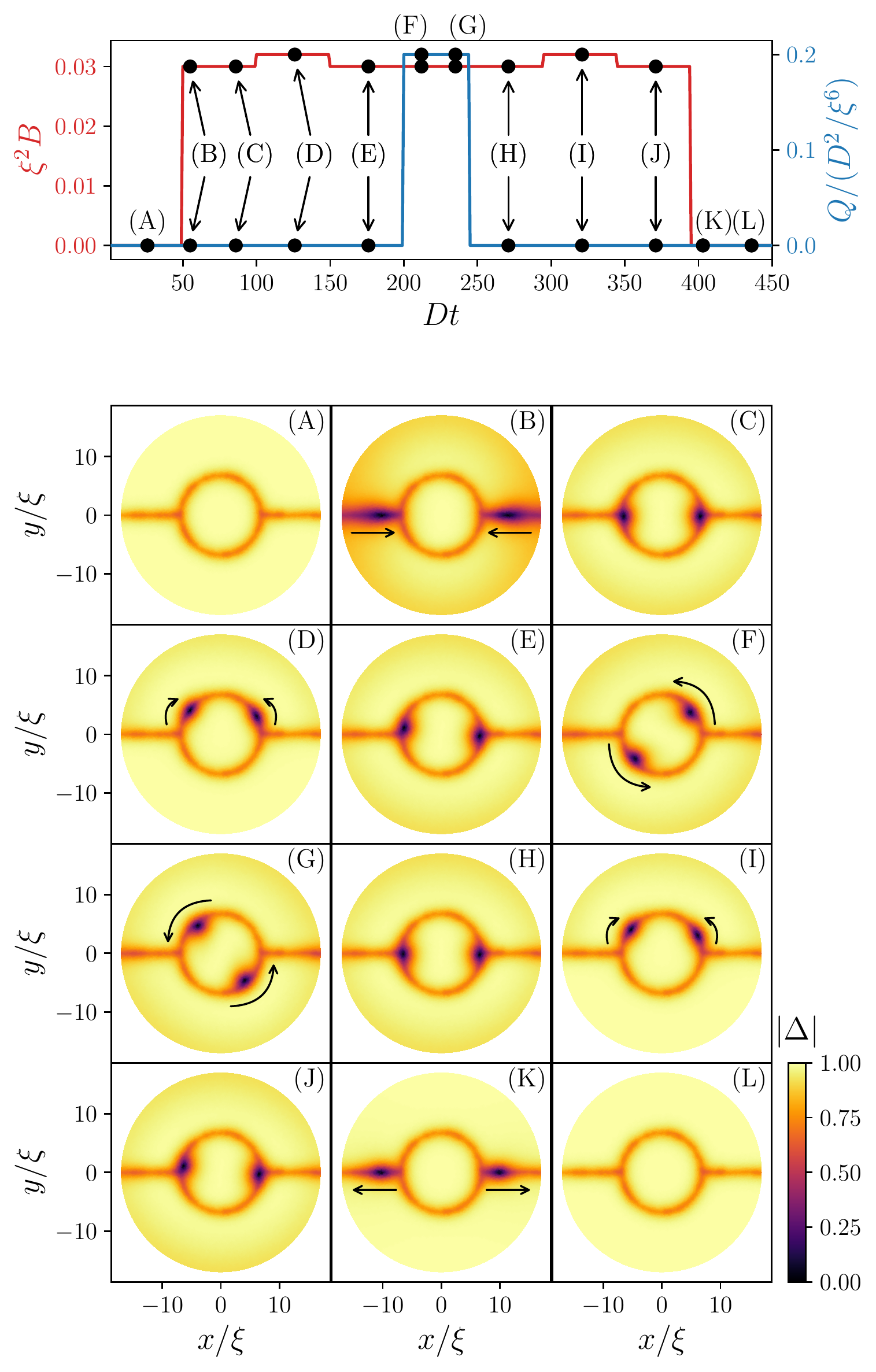}
\caption{Upper panel shows an operation scheme for initialization/read-out (fusion) and exchange processes of a pair of Majorana zero modes (bound to vortices) controlled by varying the applied field $B$ (red line with left axis) and the source term $Q$ (blue line with right axis). 
Lower panel shows snapshots of the absolute value of the order parameter $|\Delta|$ for the black points (A)-(L) in the upper panel. The arrows in the lower panel indicate the evolution of vortices (guide for the eye). The corresponding video is available in the SM\cite{SM} (see also TABLE \ref{table:video} in Appendix \ref{appendix:video}).}
\label{fig:majorana}
\end{center}
\end{figure}

\subsection{Application of vortex control}
\label{sec:application_of_vortex_control}
Finally, as one of the applications for the vortex-control paradigm, we simulate a series of operations using the 2 rails set-up investigated above. It consists of initialization/read-out (fusion) and exchange processes, which are what a quantum processor with Majorana zero modes bound to vortices is supposed to do.~\footnote{Here, we focus on the simplest case of non-topological $s$-wave superconductors, however, the main conclusions of our study should carry over to the case of topological ($p-$ or $d-$wave) superconductors, where Majorana bound state resides at the core of vortices since 
the vortex-control is rather similar and not crucially dependent on the pairing-symmetry. Our goal here is to concentrate on the vortex dynamics mimicking the quantum operations with Majorana zero modes.}

In FIG. \ref{fig:majorana}, we simulate the exemplary basic operations for initialization/read-out (fusion) and exchange processes employing two vortices. Snapshots of the absolute value of the order parameter $|\Delta|$ and an operation protocol 
are shown in the lower and upper panels, respectively (see the SM video22 for the absolute value of the order parameter $|\Delta|$ and video23 for that of phase $\phi$
~\footnote{In the video for the time dependent phase map $\phi(x,y,t)$, a $2\pi$ phase winding around each vortex and the exchange of two vortices are clearly shown. During the exchange process, the phase of the inner superconductor $\phi_{\text{in}}$ in $r<R$ changes by $-2\pi$, while that of the outer superconductor $\phi_{\text{out}}$ in $r>R$ remains almost the same. In terms of Majorana operators $\gamma_{j=1,2} = \exp{(i \psi_j)} \,c + \exp{(-i \psi_j)}\, c^{\dagger}$ with $\psi_{j}= (\phi_{\text{in}}-\phi_{\text{out}})/4+ \pi(j-1)/2$ and complex fermion operators $c$ and $c^{\dagger}$,~\cite{Sunghun2015} the process results in the exchange: $\gamma_1 \rightarrow -\gamma_2,\, \gamma_2 \rightarrow \gamma_1$. 
}in the SM~\cite{SM}).
For times $t$ with $Dt<50$, the system is in equilibrium with $B=Q=0$ and there is no vortex in the system as shown in the snapshot (A) of the lower panel. At $Dt=50$, a finite magnetic field $B$ is applied to create two vortices in the superconducting thin film as shown in the snapshot (B) of the lower panel. Here, the applied value of $B$ is determined corresponding to the one which gives two vortices in the 2 rails set-up shown as the orange line in FIG. \ref{fig:comp_all_0124N}. The two vortices stay at the intersections between the Josephson junction and the rails shown in snapshot (C) since there is no driving current ($Q=0$ at this time). To initialize a pair of Majorana zero modes one can bring them close to each other having their wave functions overlapping. This corresponds to moving the two corresponding vortices close to each other which splits the energy levels of the two parity states the two Majoranas can be fused to.~\cite{Leijnse_2012} This distinction can be used to initialize the Majorana qubit (see section~\ref{sec:operation_on_majorana_qubits}) by putting each pair into a definite parity state (e.g. the ground state parity).
We demonstrate this scheme by applying an inhomogeneous magnetic field.~\cite{Suraj2020} Specifically, in our current set-up, the inhomogeneous magnetic field is implemented in such a way that $B$ becomes non-zero only on the upper half of the outer superconductor of the Josephson junction. At $Dt=100$, the inhomogeneous field is generated and the two vortices move closer to each other as shown in snapshot (D).~\footnote{Here, the initial distance of the two vortices is only moderately reduced. However, since how close they should be to induce a coupling between the Majoranas depends on specific material parameters, we only establish here the possibility of controlling the distance of the two vortices by the inhomogeneous magnetic field.} 
Once the inhomogeneous magnetic field is set back to the homogeneous one at $Dt=150$, the two vortices go back to the original stable position, namely the intersections of the Josephson junction with the rails as observed in snapshot (E). At $Dt=200$, the source current is also applied, thus the system has both a finite $B$ and $Q$, and the two vortices start to move along the Josephson junction as shown in snapshot (F) and (G). Just before the two vortices exchange their positions, the current source $Q$ is switched off (at $Dt=245$). This switch-off time is determined by how long it takes for two vortices to exchange their positions. An exact timing is not necessary here since slight deviations are corrected as each vortex is attracted towards the closest intersection of the Josephson junction with the rails (compare the SM video20, 22, 24~\cite{SM}). After $Q$ has been set to zero, the two vortices stay at the intersections of the Josephson junction with the rails, see snapshot (H).
The result of such an exchange (braiding) operation can be read-out by fusing again corresponding pairs of Majorana zero modes by bringing their vortices in close proximity again (read-out scheme).~\footnote{Braiding two Majorana zero modes cannot change the parity of their fusion state, but it can change the parity of the fusion state of other pairs of Majorana zero modes. To perform non-trivial quantum operations by braiding, one therefore needs more than one pair of Majorana zero modes, see section~\ref{sec:operation_on_majorana_qubits}} This is done by applying the inhomogeneous magnetic field at $Dt=295$, see snapshot (I). At $Dt=345$, the inhomogeneous magnetic field is switched off and the original homogeneous field is applied. Then, at $Dt=395$, the magnetic field is also turned off and the two vortices move out of the system since there is no magnetic field anymore as shown in snapshot (K). Finally, the system goes back to the original equilibrium state with $B=Q=0$, see snapshot (L). 

\begin{figure}[t!]
\begin{center} 
\includegraphics[width=0.5\textwidth, angle=-0]{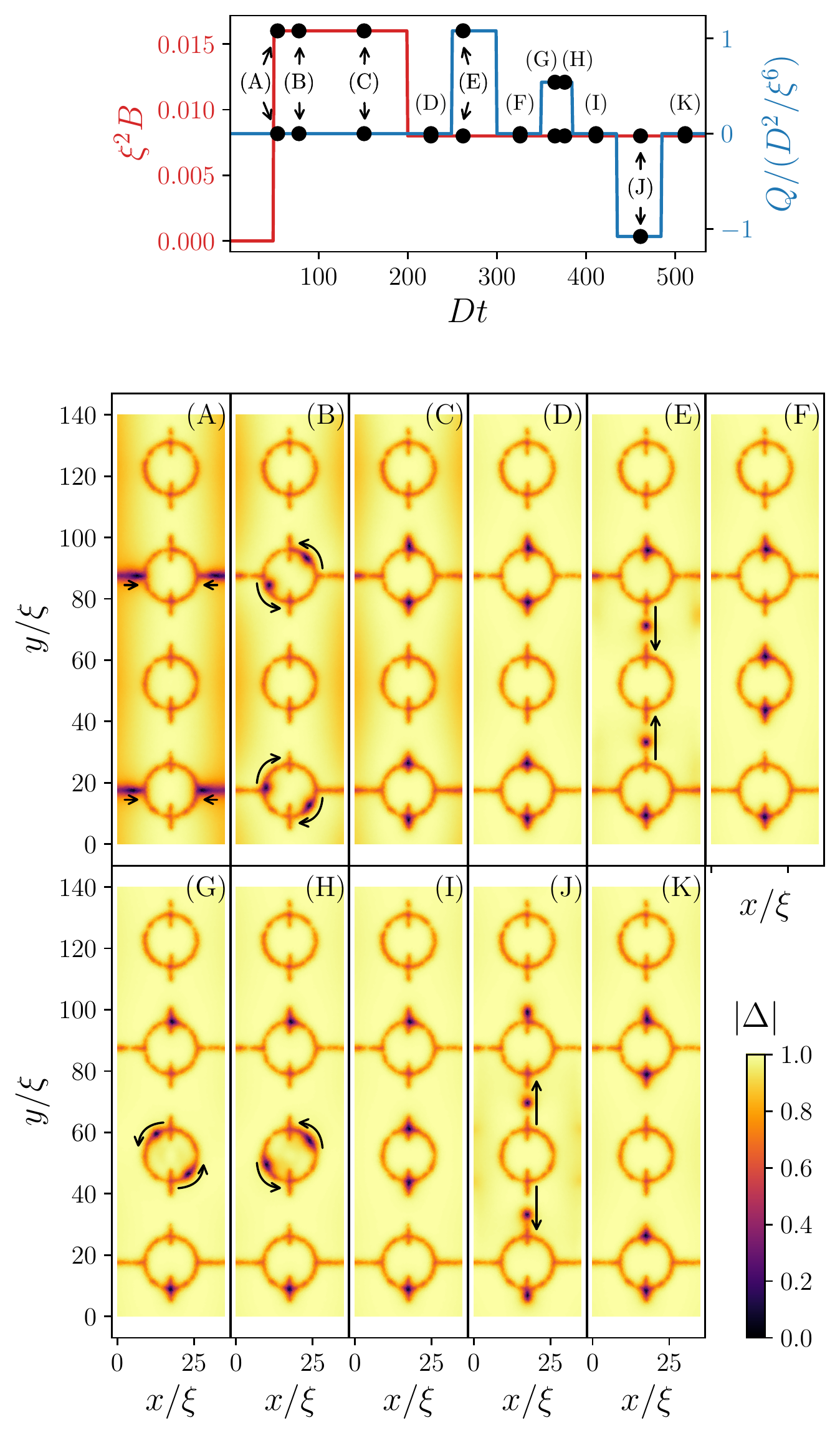}
\caption{Upper panel shows an operation scheme for an application of Majorana qubits controlled by the strength of the applied magnetic field $B$ (red line with left axis) and the source term $Q$ (blue line with right axis). Lower panel shows snapshots of the absolute value of the order parameter $|\Delta|$ for the black points (A)-(K) in the upper panel. The arrows in the lower panel indicate the evolution of vortices (guide for the eye). The corresponding video is available in the SM\cite{SM} (see also TABLE \ref{table:video} in Appendix \ref{appendix:video}).}
\label{fig:qbit}
\end{center}
\end{figure}

In order to profit from the exchange statistics of non-Abelian anyons, as we mentioned in section \ref{introduction}, we need at least four Majorana zero modes. Here, we show results of a simulation for controlling four vortices using three Corbino geometry Josephson junctions with additional rails. The representative result is shown in FIG. \ref{fig:qbit}, where snapshots of the absolute value of the order parameter $|\Delta|$ and a protocol for 
the braiding operation are shown in the lower and upper panels, respectively (see the SM video26~\cite{SM}). Unlike the disk geometry we use above, here, we employ a rectangular geometry with periodic boundary conditions along the $y$ direction, e.g. $\Delta(y+L_y)=\Delta(y)$, where $L_y$ is the length of the sample in y-direction. The length of the finite strip geometry in $x$ and $y$ directions are set to be $L_x=35 \xi (\equiv L)$ and $L_y=4L$, respectively. We choose the vector potential $\bm{A}=(0, B(x-L/2), 0)$, which satisfies the boundary conditions of the current set-up. Four Corbino geometry Josephson junctions are prepared on the sample (we call these four Corbino geometries first, second, third, and fourth with respect to $y=0$). The radius and width of the ring-shaped normal metallic regions are $8.5 \xi (\equiv R)$ and $0.5\xi$, respectively. Even though the fourth Corbino geometry is not used in this simulation, we introduce it because together with periodic boundary conditions, the current set-up essentially demonstrates a register of qubits, which is more practical since multiple qubits are required for the operation of a quantum processor. Thus, there is no reason to believe, that the general results presented here cannot be incorporated and extended into columns of such interconnected quasi-1D Corbino geometry sequences on 2D superconducting materials.

Now, let us look into a result of the simulation. We begin with the equilibrium situation as in FIG. \ref{fig:majorana}, namely $B=Q=0$. At $Dt=50$, we switch on the magnetic field $B$ to create four vortices on the superconducting thin film as shown in the snapshot (A) of the lower panel, where two vortices nucleate in the first and third Corbino geometries via their rails. After the nucleation of vortices, these four vortices spontaneously slide to $x=L/2$ through Josephson junctions since there is open space between the first and third Corbino geometry Josephson junctions, and the two vortices in the same Josephson junctions repel each other This dynamics is shown in the snapshot (B). After the two vortices achieve $x=L/2$ in each Corbino geometry, they stay at the intersections between the Josephson junction and the non-connected rails at $x=L/2$ as shown in the snapshot (C). 
Next, one of the two vortices in each junction is sent to the second Corbino geometry Josephson junction. For this purpose, we induce current flow in $x-$direction around the close vicinity of vortices 
in such a way that two vortices located around $y=L/2+R$ and $y=2.5L-R$ are pushed toward the second Corbino geometry Josephson junction due to the Lorentz force. Thus, following the same implementation for creating source current as demonstrated in the disk geometry, we introduce an external current created by a source and drain of particles 
whose strength is denoted by $Q$, on both sides of the open boundaries  i.e. $x=0$ and $x=L$ with width $\xi$. 
These sections are extended between $y=L/2$ (middle point of the first Corbino geometry) and $y=1.5L-R$ (the lower intersection between the second Corbino geometry and non-connected rail) and between $y=1.5L+R$ (the upper intersection between the second Corbino geometry and non-connected rail) and $y=2.5L$ (middle point of the third Corbino geometry).

Before switching on the source current, we reduce the strength of the magnetic field to ensure that additional vortices do not nucleate in the system. Here, we choose the magnetic field to be half of the original value which does not change the number of vortices in the system as shown in the snapshot (D).~\footnote{According to our simulation, if we simply use the original magnetic field value, additional vortices nucleate when one of the two vortices in each junctions is sent to the second Corbino geometry Josephson junction.} This also shows that vortices are so strongly trapped at the intersection that they do not leave the system. Then, the externally applied source current we defined above is switched on and one of the two vortices in each junction is sent to the second Corbino geometry Josephson junction as shown in the snapshot (E). At $Dt=300$, the source current is switched off and the two transported vortices are trapped at an intersection between the Josephson junction and non-connected rails of the second Corbino geometry. Then, the braiding operation is performed as demonstrated in the case of the single Corbino geometry as shown in FIG. \ref{fig:majorana}. Thus, the source and drain are created at the outside and inside of the second Corbino geometry whose shapes are a ring with radius/width $16.75\xi/0.5\xi$ and a disk with radius $2.9\xi$, respectively. (The structure of the source and drain sections are same as for the case of the disk geometry used before. See also Appendix \ref{appendix:source} for details
). The corresponding snapshots are shown in (G),(H). After the externally applied source current is turned off, the two vortices used for the braiding operation stay at the intersection between the Josephson junction and the non-connected rails as shown in snapshot (I). Finally, these exchanged vortices are sent back to the first and third Corbino geometries, respectively by the externally applied source current from the edge as used before which is shown in snapshot (J). This time, the direction of current flow should be opposite to the case of (E) to induce the reverse vortex motion. In the numerical simulation, this can be done simply by switching the sign of the charge in the source and drain sections compared to the case of (E)($\pm Q \rightarrow \mp Q$). After this, the externally sourced current is set to zero and all four vortices are trapped at the intersection between the Josephson junction and the non-connected rails. 

\subsection{Operations on Majorana qubits} 
\label{sec:operation_on_majorana_qubits}
The proposed operation schemes for vortices can be implemented to perform operations on Majorana zero modes by proximity-inducing the order parameter of a $s$-wave superconductor into the surface of a three-dimensional topological insulator (TI) where spin-helical electrons reside. In the TI-based Corbino Josephson junctions, an effective $p$-wave pairing is produced and each vortex binds a single Majorana zero mode, which we call Majorana vortex. In this section, we interpret the simulation of vortex dynamics shown in section \ref{sec:application_of_vortex_control} in terms of Majorana vortices, 
and illustrate how the operations shown in FIGs. \ref{fig:majorana} and \ref{fig:qbit} act on a Majorana qubit. 

\begin{figure}[t!]
\begin{center} 
\includegraphics[width=0.5\textwidth, angle=-0]{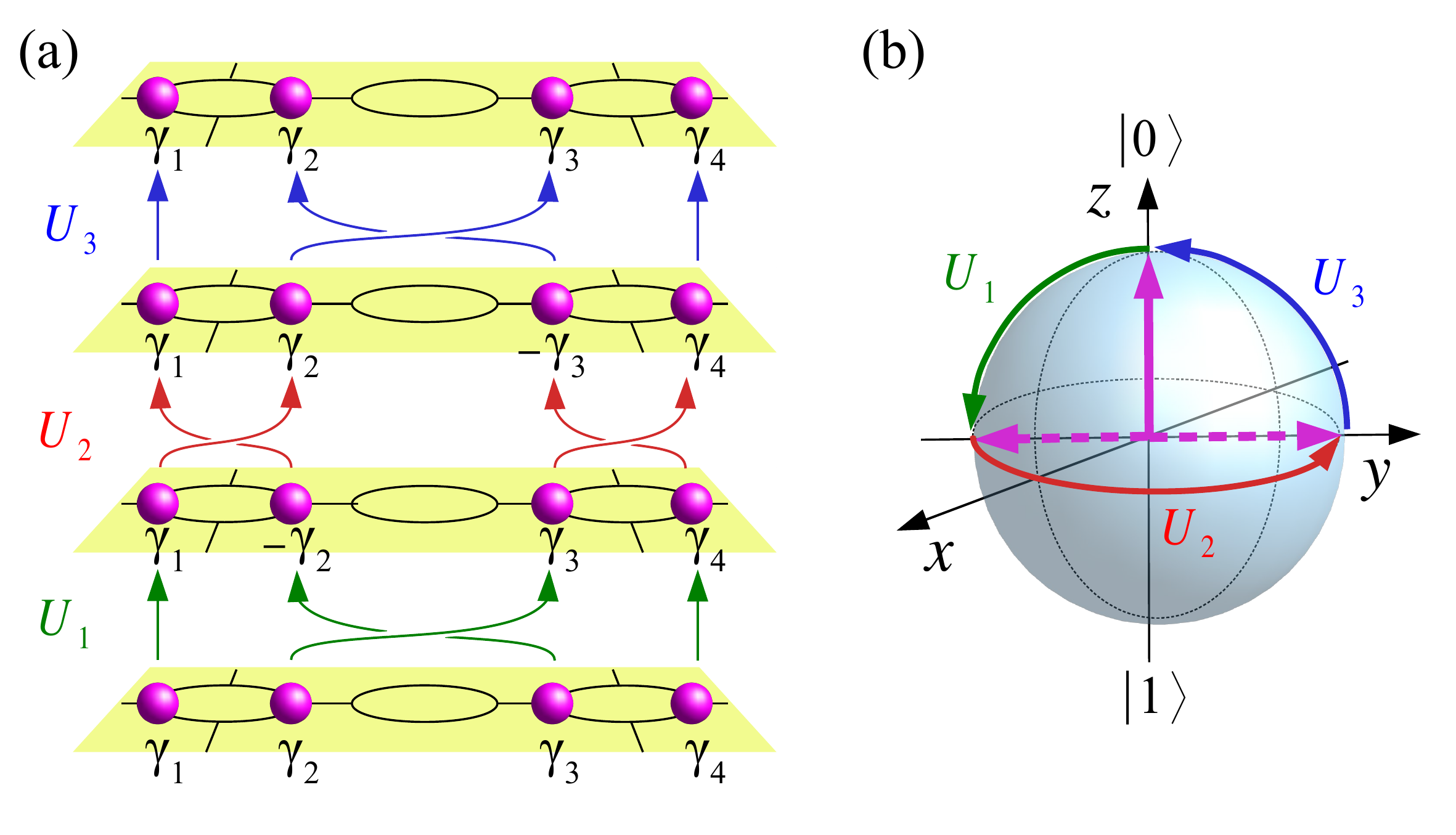}
\caption{Schematic illustration of Majorana vortices $\gamma_i$, denoted by magenta balls, in three Corbino Josephson junctions connected by rails. (a) A sequence of operations by adopting the simulations in FIGs. \ref{fig:majorana} and \ref{fig:qbit}. (b) Bloch sphere representation of the evolution of Majorana qubit, indicataed by arrows in magenta, corresponding to the process in (a).}
\label{MQubit}
\end{center}
\end{figure}

We consider four Majorana vortices $\gamma_i$ with $i \in \{1,2,3,4\}$ in three Corbino Josephson junctions connected by rails, see (a) of FIG. \ref{MQubit}. We prepare the configuration where one pair of Majorana vortices is in the left junction and the other is in the right junction, and assume that the total number of fermions in the system is conserved. They define a single Majorana qubit, 
\begin{equation}
|0\rangle \equiv |0_{12}0_{34}\rangle, \hspace{8pt} |1\rangle \equiv |1_{12}1_{34}\rangle, 
\end{equation}
where $|1_{12}1_{34}\rangle = c^{\dagger}_{12} c^{\dagger}_{34}|0_{12}0_{34}\rangle$ with complex fermion operators, $c^{\dagger}_{12}= (\gamma_1-i\gamma_2)/2$ and $c^{\dagger}_{34}= (\gamma_3-i\gamma_4)/2$, built from the Majorana modes; note that one can also define a qubit with odd fermion number states, $|1_{12}0_{34}\rangle \equiv c^{\dagger}_{12}|0_{12}0_{34}\rangle$ and $|0_{12}1_{34}\rangle \equiv c^{\dagger}_{34}|0_{12}0_{34}\rangle$. Here, the eigenvalues $n_i=0, 1$ of the occupation number operators $c^{\dagger}_i c_i$ corresponding to the empty and filled states of the fermions, respectively, reflect the fusion of two Majorana vortices. In this basis, the braiding operations depicted in (a) of FIG. \ref{MQubit} are represented as 
\begin{equation}
U_1=U_3= \frac{1}{\sqrt{2}}\left(1- i \sigma_x \right), \hspace{8pt} U_2=-i \sigma_z,
\end{equation}    
where $\sigma_i$ are the Pauli matrices in the qubit space. Here, $U_2$ describes the braiding of a pair of Majorana in each junction by adjusting the source currents shown in FIG. \ref{fig:majorana} and $U_{1,3}$ are that of Majorana vortices from different junctions based on the simulation in FIG. \ref{fig:qbit}. A sequence of the operations $U_3 U_2 U_1$ acting on the Majorana qubit on a Bloch sphere is illustrated in (b) of FIG. \ref{MQubit}. The unitary operations can be used to produce the Pauli gates up to a phase factor,
\begin{equation}
U^2_1= -i \sigma_x, \hspace{5pt} U_2 U^2_1= -i \sigma_y, \hspace{5pt} U_2= -i \sigma_z.
\end{equation} 
The read-out of the quantum operation can be done in the computational basis $|0_{12}0_{34}\rangle, |1_{12}1_{34}\rangle$ (or $|1_{12}0_{34}\rangle, |0_{12}1_{34}\rangle$) by bringing two vortices close together (as shown in (D) of FIG. \ref{fig:majorana}), e.g. by controlling a local magnetic field, which lifts the degeneracy between the empty and filled fermion states.

\section{Conclusion}
\label{conclusion}
We study dynamics of vortex motion in Corbino geometry superconducting-normal-superconducting Josephson junctions on a superconducting thin film in the framework of the time-dependent Ginzburg Landau theory. As expected, we observe that vortices move along the Josephson junction due to the Lorentz force created by an externally sourced current flowing from the boundary of the circle to the center
, and this circular motion of vortices on the Josephson junction can be captured through the jumps between resistance plateaus. Results of our simulation for the conventional Corbino geometry indicate difficulties of controlling vortices, namely the one by one increase in the number of vortices with increasing an applied magnetic field, unpredictable nucleation points of vortices, and trapping all the vortices on the circular Josephson junction. This could explain the absence of quantization in recent experiments.~\cite{Matsuo2020} By installing additional railings to the Corbino geometry set-up, these three issues can be resolved and, specifically, the 4 rails set-up ensures that all vortices stay on the Josephson junction after the one by one vortex nucleation within certain externally sourced current and applied magnetic field regimes.

Further analysis for the defect edge and the 4 rails set-ups by sweeping through the source current and the applied magnetic field reveals that a larger source current yields a larger resistance due to an increase in the speed of the vortex motion on the Josephson junction, and the 4 rails set-up with up to moderate strength of the source currents can prevent the first nucleated vortex from going into the center of the superconductor sample rather than being trapped on the Josephson junction.


As a prospective application of the vortex control in Corbino geometry Josephson junctions, we demonstrated the braiding (spatial exchange) of two vortices that can be brought close together (fused) before and after the exchange by tuning the external magnetic field and the source current. These basic control elements would allow to read-in, operate and read-out qubits based on Majorana zero modes.
Isolated Majorana zero modes are non-abelian anyons (Ising anyons) that are confined at the center of vortices in spinless $p-$wave superconductors. We also proposed a (scalable) platform of interconnected Corbino rings where four vortices can be manipulated in a way that corresponds to non-abelian state operations for Majorana zero modes bound to the vortices.

\begin{acknowledgments} 
\noindent \textit{Acknowledgments}.---
This work was supported by the Deutsche
Forschungsgemeinschaft (DFG, German Research Foundation) via RTG 1995 and Germany’s Excellence Strategy - Cluster of Excellence Matter and Light for Quantum Computing (ML4Q) EXC 2004/1 - 390534769 as well as via Germany’s Excellence Strategy-EXC-2123 QuantumFrontiers-390837967. Simulations were performed with computing resources granted by RWTH Aachen University under project rwth0601 and rwth0507.
\end{acknowledgments}

\appendix
\section{Structure of source and drain}
\label{appendix:source}
\begin{figure}[htbp!]
\begin{center} 
\includegraphics[width=0.49\textwidth, angle=-0]{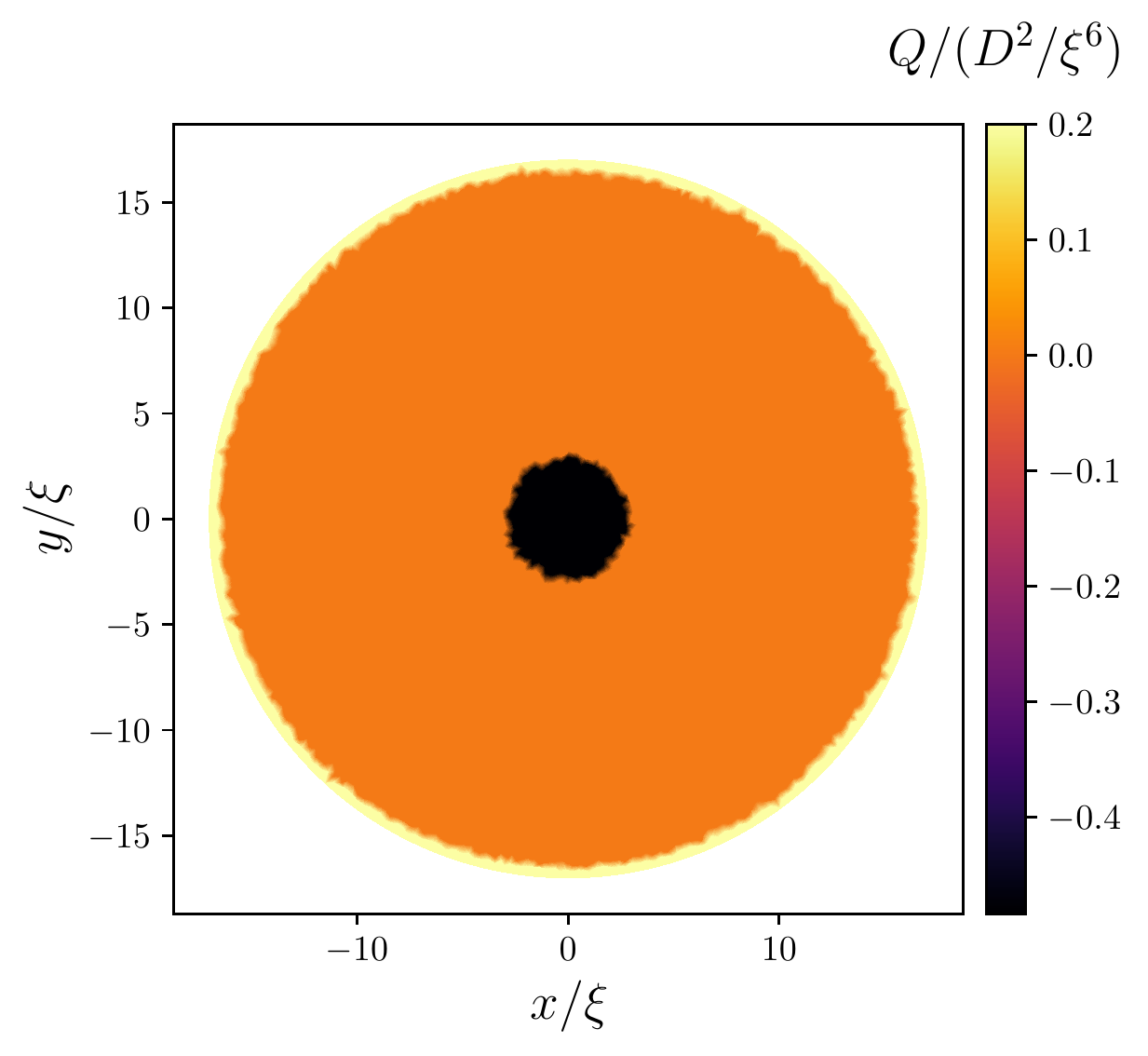}
\caption{Structure of the source and drain creating an externally sourced current whose strength is denoted by $Q$ in Eq. \eqref{continuity}. The source and drain sections are in the vicinity of the open boundary of the superconducting disk and in its center, respectively. The width of the source section is $0.5\xi$. The drain section is implemented as a disk whose radius is $2.9\xi$. This set-up is used for all simulations except for the one in FIG. \ref{fig:qbit}. 
Here, we show the structure with $Q/(D^2/\xi^6)=0.2$ as an example.}  
\label{fig:ju}
\end{center}
\end{figure}
The structure of the source and drain is shown in FIG. \ref{fig:ju}. This implementation induces homogeneous (except for the source and drain sections) current flow from the circle boundary to the center of the sample. This structure is employed for all simulations except for the one in FIG. \ref{fig:qbit}.  

\section{Distribution of averaged scalar potential $\Bar{\Theta}$ }
\label{appendix:scalar}
\begin{figure}[htbp!]
\begin{center} 
\includegraphics[width=0.49\textwidth, angle=-0]{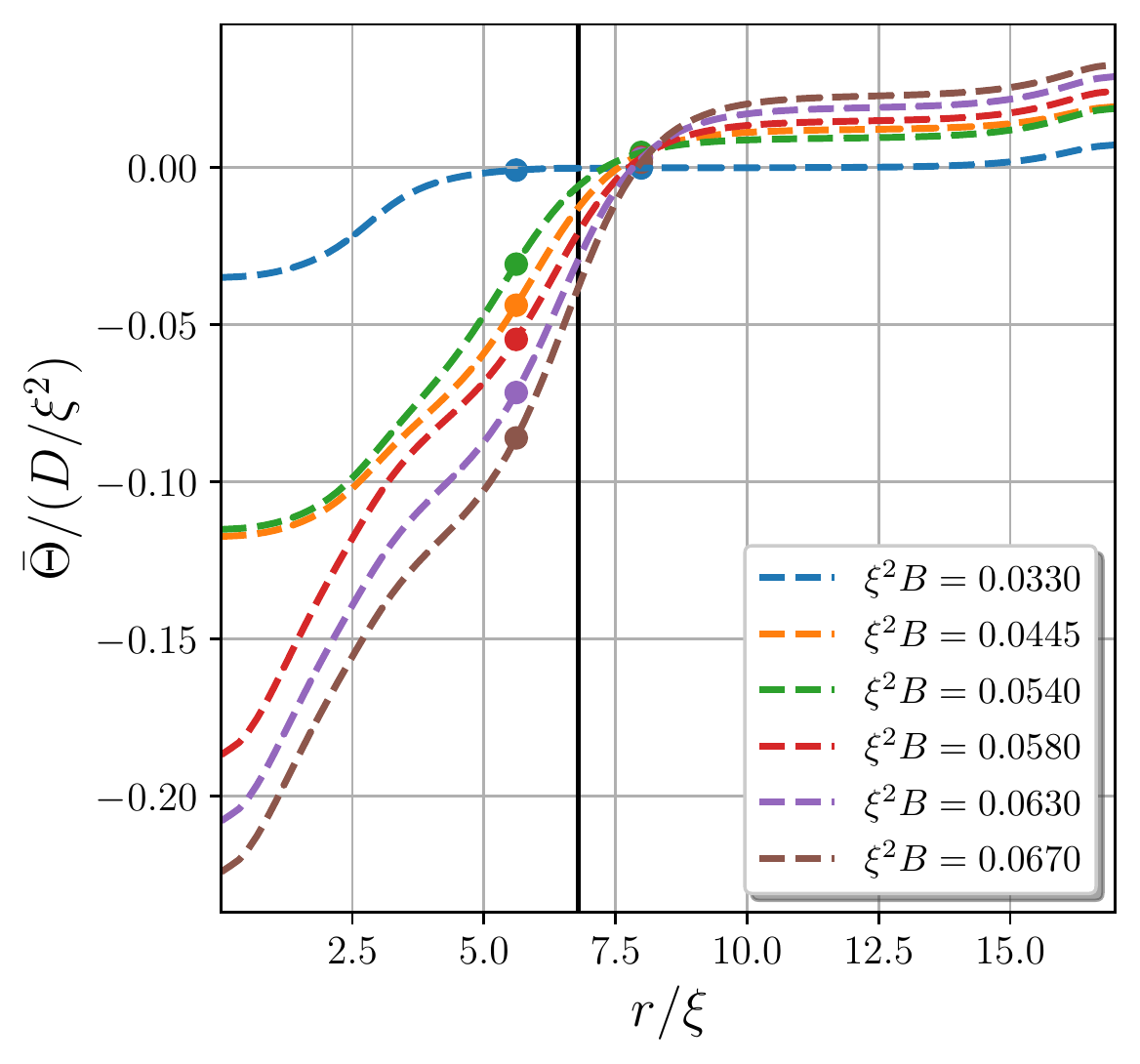}
\caption{Distribution of the averaged scalar potential $\Bar{\Theta}$ for the normal Corbino set-up used for making plots in panel (a),(b) of FIG. \ref{fig:corbino_snap_resistance}. As explained in FIG. \ref{fig:corbino_snap_resistance}, the resistance $\Delta \bar \Theta$ are calculated by taking differences of the averaged scalar potential $\Bar{\Theta}$ between two nearby points across the Josephson junction, denoted by $\Bar{\Theta}_{c_{+/-}}$ and source and drain points, denoted by $\Bar{\Theta}_{s/d}$. Specifically, the two dot points in this figure are used for the former while the origin and the end of each line are used for the latter. The vertical black line corresponds to the middle of the Josephson junction. $r$ is the radial coordinate from the center to the open boundary of the sample. The chosen parameters $B\xi^2$ correspond to the values of the black dot points denoted by (A)-(F) for the normal Corbino set-up in FIG. \ref{fig:corbino_snap_resistance}.}
\label{fig:theta_normal}
\end{center}
\end{figure}
\begin{figure}[htbp!]
\begin{center} 
\includegraphics[width=0.49\textwidth, angle=-0]{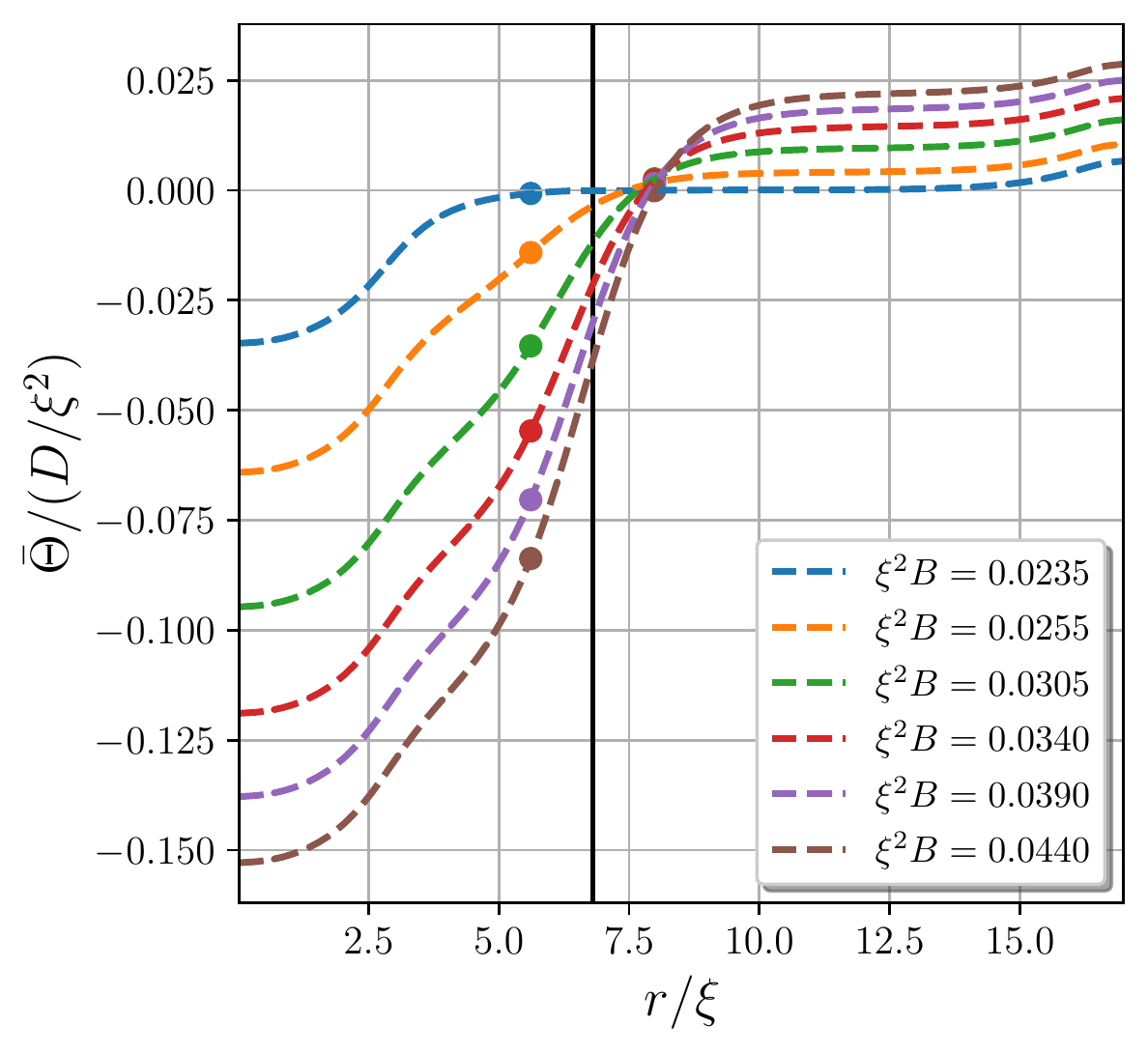}
\caption{Distribution of the averaged scalar potential $\Bar{\Theta}$ for the 4 rails set-ups used for making plots in panel (a),(b) of FIG. \ref{fig:comp_all_0124N}. As explained in FIG. \ref{fig:comp_all_0124N}, the resistance $\Delta \bar \Theta$ is calculated by taking differences of the averaged scalar potential $\Bar{\Theta}$ between two nearby points across the Josephson junction, denoted by $\Bar{\Theta}_{c_{+/-}}$ and source and drain points, denoted by $\Bar{\Theta}_{s/d}$. Specifically, the two dot points in this figure are used for the former while the origin and the end of each line are used for the latter. The vertical black line corresponds to the middle of the Josephson junction. $r$ is the radial coordinate from the center to the open boundary of the sample. The chosen parameters $B\xi^2$ correspond to the values of the black dot points denoted by (A)-(F) for the 4 rail set-up in FIG. \ref{fig:comp_all_0124N}.}
\label{fig:theta_4rails}
\end{center}
\end{figure}
In this section, we show the distributions of the averaged scalar potential $\Bar{\Theta}$ over both space and time for the normal Corbino and 4 rails set-ups shown in FIG. \ref{fig:theta_normal} and FIG. \ref{fig:theta_4rails}, respectively. As is mentioned in the main text, when a vortex resides in the center of the sample, the distributions of the averaged scalar potential $\Bar{\Theta}$ at that vicinity (aka around the center of the drain section) shows the deep downward slope toward the center, which is clearly seen by comparing the blue, green, orange and red, purple, brown curves in FIG. \ref{fig:theta_normal}. They correspond to the snapshots (A)-(C) and (D)-(F) in the lower panel of FIG. \ref{fig:corbino_snap_resistance}, respectively.     

\section{Short explanation of the videos}
\label{appendix:video}
\begin{table}[t!]
\begin{center}
\caption{Short explanation of the videos in the SM~\cite{SM}. This table presents an overview of the video in the SM and links them to all the figures in the main text.}
{
\begin{tabular}{ccccccc}\hline\hline
video  && set-up & quantity & $Q/(D^2/\xi^6)$ & FIGs & $T_{\text{on}/\text{off}}$~\footnote{The time of switching on/off the source current for exchanging vortices. The unit is $D^{-1}$. This column is applicable only for the simulations involving the vortex exchange.} \\ \hline
1    && normal & $|\Delta|$ & 0.20 & \ref{fig:corbino_snap_resistance}   \\
2    && defect edge & $|\Delta|$ & 0.20 & \ref{fig:comp_all_0124N}, \ref{fig:defect1_snap}, \ref{fig:phase_defect}  \\
3    && 1rail & $|\Delta|$ & 0.20 & \ref{fig:comp_all_0124N}, \ref{fig:defect1_snap}   \\ 
4    && 2rail & $|\Delta|$ & 0.20 & \ref{fig:comp_all_0124N}, \ref{fig:24rail_snap}   \\
5    && 4rail & $|\Delta|$ & 0.20 & \ref{fig:comp_all_0124N}, \ref{fig:24rail_snap}, \ref{fig:phase_4rail}   \\
6    && defect edge & $|\Delta|$ & 0.05 & \ref{fig:phase_defect}   \\
7    && defect edge & $|\Delta|$ & 0.10 & \ref{fig:phase_defect}   \\
8    && defect edge & $|\Delta|$ & 0.15 & \ref{fig:phase_defect}   \\
9    && defect edge & $|\Delta|$ & 0.25 & \ref{fig:phase_defect}   \\
10    && defect edge & $|\Delta|$ & 0.30 &\ref{fig:phase_defect}   \\
11    && defect edge & $|\Delta|$ & 0.35 & \ref{fig:phase_defect}   \\
12    && defect edge & $|\Delta|$ & 0.40 & \ref{fig:phase_defect}   \\
13    && 4rail & $|\Delta|$ & 0.05 & \ref{fig:phase_4rail}   \\
14    && 4rail & $|\Delta|$ & 0.10 & \ref{fig:phase_4rail}   \\
15    && 4rail & $|\Delta|$ & 0.15 & \ref{fig:phase_4rail}   \\
16    && 4rail & $|\Delta|$ & 0.25 & \ref{fig:phase_4rail}   \\
17    && 4rail & $|\Delta|$ & 0.30 & \ref{fig:phase_4rail}   \\
18    && 4rail & $|\Delta|$ & 0.35 & \ref{fig:phase_4rail}   \\
19    && 4rail & $|\Delta|$ & 0.40 & \ref{fig:phase_4rail}   \\ 
20    && 2rail & $|\Delta|$ & 0.20 & \ref{fig:majorana} & 200/240  \\
21    && 2rail & $\phi$ & 0.20 & \ref{fig:majorana} & 200/240  \\
22    && 2rail & $|\Delta|$ & 0.20 & \ref{fig:majorana} & 200/245  \\
23    && 2rail & $\phi$ & 0.20 & \ref{fig:majorana} & 200/245  \\
24    && 2rail & $|\Delta|$ & 0.20 & \ref{fig:majorana} & 200/250  \\
25    && 2rail & $\phi$ & 0.20 & \ref{fig:majorana} & 200/250  \\ 
26    && special~\footnote{A special set-up used for the qubit simulation in section~\ref{sec:operation_on_majorana_qubits}} & $|\Delta|$ &0.54~\footnote{The $Q/(D^2/\xi^6)$ value for exchanging vortices} & \ref{fig:qbit} & 375/420 \\ \hline
\end{tabular}
}
\label{table:video}
\end{center}
\end{table}
In this section, we illustrate an overview of the videos found in the SM~\cite{SM} and connect them to all the figures in the main text. This is summarized in TABLE \ref{table:video}.

\bibliography{prb}%

\end{document}